\documentstyle[11pt,aaspp]{article}

\received{22 September 1997}
\slugcomment{To appear in A. J., February 1998}

\def\ltsim{\raisebox{-.5ex}{$\;\stackrel{<}{\sim}\;$}}
\def\gtsim{\raisebox{-.5ex}{$\;\stackrel{>}{\sim}\;$}}
\def\cc{\ifmmode {\rm cm}^{-3} \else cm$^{-3}$\fi}
\def\cl{\ifmmode {\rm cm}^{-2} \else cm$^{-2}$\fi}
\newcommand{\rf}{\reference}
\newcommand {\etal}{et~al.}
\newcommand {\kms}{{$\rm km\,s^{-1}$}}

\newcommand {\lya}{Ly$\alpha$}
\newcommand {\hi}{H$_0$}
\newcommand {\h}{H~{\sc i}}
\newcommand {\he}{He~{\sc ii}}
\newcommand {\hep}{He$^+$}
\newcommand {\HI}{\rm H{\scriptscriptstyle I}}
\newcommand {\HeII}{\rm He{\scriptscriptstyle II}}
\newcommand{\me}{Miralda-Escud\'e}
\newcommand{\gp}{Gunn-Peterson}

\begin{document}


\title{The He {\normalsize\bf II} Opacity of the Lyman $\alpha$ Forest
and \\the Intergalactic Medium}

\author{Wei Zheng, Arthur F. Davidsen, and Gerard A. Kriss}
\affil{Center for Astrophysical Sciences\\Johns Hopkins University,
Baltimore, MD 21218\\zheng@pha.jhu.edu afd@pha.jhu.edu gak@pha.jhu.edu}
\begin{abstract}
\rightskip 0pt plus 2pt

Primordial baryonic matter in the vast intergalactic space may be traced
with
\lya\ resonance absorption by neutral hydrogen and singly ionized helium.
The wavelength-averaged \he\ opacity shortward of 304(1+z)~\AA, as
measured by
low-resolution UV spectroscopy, is at least 4.5 times the \h\ opacity
shortward of 1216(1+z)~\AA. While a part of the \he\ opacity arises from
the
intergalactic regions that produce the known \lya\ forest, it has
been argued whether the \he\ opacity may be {\em entirely} attributable to
these observed lines.
Based on the empirical formulas governing the distribution of \lya\ forest
absorption, we use a Monte-Carlo technique to calculate the average \he\
optical depth produced by these forest lines.

The \he\ counterparts of the \lya\ forest lines are highly saturated, and
hence
their contributions to the observed opacity is limited.
Assuming a \hep/\hi\ population ratio of 100 and that the
power-law distribution ${dn/dN} \propto N^{-1.5}$ can be extended
to a neutral hydrogen column density of $N_{H I} = 2 \times
10^{12}$~\cl, the contribution from these forest lines may account
for a \he\ opacity that is $\ltsim 3$ times the \h\ opacity.
Our simulated \he\ spectrum of the quasar Q0302-003, based on the fitted
lines in the high-resolution Keck spectrum, yields a forest
optical depth of $\sim 0.9$, less than  half of the observed \he\
opacity. Therefore, a substantial contribution to \he\ absorption
arises from extremely tenuous regions of intergalactic gas that are
beyond the observational limits for \h\ absorption. \he\ spectra
at higher resolution are a sensitive tool to explore the properties of
these
small-scale fluctuations that fill $\sim 80\%$ of the intergalactic space
and contain a significant part of the baryonic matter in the early
universe.
\end{abstract}

\newpage

\section{INTRODUCTION}

Shortly after the discovery of quasars, it was found that the flux
shortward of the redshifted \lya\ emission at (1+z) 1216 \AA\ is
depressed. Gunn \& Peterson (1965) suggested that this may be due to the
accumulated
\lya\ absorption by a uniform, diffuse intergalactic medium (IGM) at
high redshifts. Optical spectroscopy at improved spectral resolution
(Lynds 1971; Sargent et al. 1980) found that the absorption arises from
numerous narrow features referred to as ``the \lya\ forest''.
The lack of a smooth depression in flux
(the \gp\ effect, see Steidel \& Sargent 1987 and references therein)
has led to the conclusion that the IGM is highly
ionized by a metagalactic UV radiation field in the early
universe (Fields 1972; Bechtold et al. 1987; \me\ \& Ostriker 1990).
Premordial intergalactic helium should also
produce absorption features, notably \he\ \lya.
Recent theoretical models of the IGM evolution  (Cen \etal\ 1994;
Zhang, Anninos, \& Norman 1995; \me\ \etal\ 1996; Hernquist et al. 1996;
Bi \& Davidsen 1997) suggest that there is no distinct boundary
between a diffuse component of the IGM and discrete forest clouds.
The evolution of the universe leads to density fluctuations on all scales.
Evolving structures produce absorption features with a column-density
distribution similar to that observed in the \lya\ forest.
The features of the lowest column density arise in underdense portions of
the IGM and contribute most to the \he\ opacity.
In conventional terms, forest lines are referred to as those that can be
identified with optical spectroscopy. Most of them have column densities
of
$10^{13}$ \cl\ or higher, and the Keck HIRES spectra have identified
absorption
features as weak as  $N = 2 \times 10^{12}$ \cl\ (Hu et al. 1995).

If the UV background radiation is derived mainly from quasar radiation,
its EUV photons are able to ionize \hep, and hence the
intergalactic helium is likely to be primarily in the form of He$^{++}$
(Madau 1992).
However, because of the high ionization level for \hep, the \hep\
population
in the IGM would be significantly higher than that of \hi\
\markcite{(Sargent et al. 1980; \me\ \& Ostriker 1990)} and may hence be
more easily detected. \he\ Ly$\alpha$ absorption is within the
spectroscopic
capability of the Hubble Space Telescope ({\it HST}) for quasars with $\rm
z >
3$ and of the Hopkins Ultraviolet Telescope (HUT) for quasars of $\rm z >
2$.

\markcite{Jakobsen et al. (1994)} reported a cutoff in flux
shortward of redshifted \he\ \lya\ emission in the {\it HST}
Faint Object Camera (FOC) prism spectrum of the quasar Q0302-003 ($\rm
z_{em}
= 3.286$),
with optical depth $ \tau > 1.7$ at 90\% confidence level.
Follow-up {\it HST HRS}
observations (Hogan, Anderson \& Rugers 1997) find that the average
\he\ optical depth  $\tau_{He II} = 2.0^{+1.0}_{-0.5}$
(95\% confidence level) at redshift $\rm z \approx 3.15$.
The spectrum of the quasar HS1700+64 ($\rm z_{em} = 2.743$) obtained with
HUT
had previously provided a measurement of the average optical depth $\tau =
1.00 \pm 0.07$
at $\rm z \approx 2.4$ (Davidsen, Kriss, \& Zheng 1996). The long-sought
\he\
absorption confirms primordial intergalactic helium as predicted by the
Big Bang theory and provides insights into the ionization state and
density of the IGM.

Neither of these observations, however, can resolve \he\ \lya\ forest
lines,
hence the observed opacity in the spectral region shortward of
304(1+z)~\AA\
is a blend of absorption features arising from regions of various density.
The regions with very small-scale fluctuations in the IGM produce
absorption features that are below the observational limit, i.e.  they
have
with the neutral hydrogen column density $N_{H I} < 2 \times 10^{12}$ \cl\
($N$ hereafter). By precisely modeling the expected contribution of the
observed \he\ \lya\ forest features based on measurements of the \h\ \lya\
forest, one can hope to disentangle the contributions of the different
regions of the IGM above and below the threshold column density $N = 2
\times 10^{12}$ \cl.

Several techniques have been used to estimate the likely
contribution of the observed forest lines to the \he\ opacity.
Jakobsen \etal\ (1994), Madau \& Meiksin (1994), and
Giroux, Fardal, \& Shull (1995) all used the formula
\begin{equation}
\tau ( z ) = {{dn} \over {dz}} {{< W_\lambda >} \over {\lambda}} ( 1 + z )
\end{equation}
where $dn / dz$ is the number of lines per unit redshift and $< W_\lambda
>$
is the mean rest-frame equivalent width of \he\ \lya\ lines averaged over
the assumed distribution in column density (M\o ller \& Jakobsen 1990).
This in turn is based on the optical depth for a random distribution
of absorbing clouds as described by Paresce, McKee, \& Bowyer (1980).
Jakobsen \etal\ (1994) concluded that the \he\ forest may not account for
the observed \he\ opacity in their spectrum of Q0302-003
unless there is a significant population of clouds with $N < 10^{13}$~\cl.
Madau \& Meiksin (1994) and Giroux \etal\ (1995) explored further
how the forest opacity is influenced by the helium to hydrogen
velocity-dispersion ratio ($\xi = b_{He}/b_{H}$). Starting from a
relatively
high estimate of the \h\ opacity in the \lya\ forest toward
Q0302-003, they found that all the observed opacity can be attributed
to forest lines with $N > 10^{12}$~\cl.
A key assumption in the method upon which all these studies are based,
however, is that the absorption at any wavelength is due to only a single
cloud. When the minimum column density is extended much below $10^{13}~\rm
cm^{-2}$, observed lines begin to overlap significantly, and this
approximation breaks down.

Songaila, Hu, \& Cowie (1995) used two different approaches.
The first directly used their high resolution, high S/N ratio
spectrum of the \lya\ forest in Q0302-003.
For a normalized spectrum of the \h\ forest with intensity given by
$S (4 \lambda )$, the normalized spectrum of the \he\ forest is given by
$ S(\lambda) = [S ( 4 \lambda )]^{\eta \over 4}$, where $\eta$ is the
population ratio of \hep\ to \hi.
The principle drawback of this elegant and rather straightforward method
is
that it is extremely sensitive to the S/N level and to the fitted
continuum level.
Even at S/N = 50, noise in the data is amplified into features
in the \he\ forest with optical depth $\sim 0.4$ if $\eta = 80$.
With this approach, metal-line systems in the \h\ forest are also
incorrectly
incorporated into the predicted \he\ opacity.

Their second approach used the formula developed by \me\ (1993) to
describe
the relative opacities in the \h\ and \he\ forest:
\begin{equation}
\tau_{\HeII} = \tau_{\HI} ( {{\eta} \over 4} )^{\beta - 1} \xi^{2 -
\beta},
\end{equation}
where $\beta$ is the power law index of the \lya\ forest column density
distribution.
This formula is only valid, however, if the distribution includes lines
that are optically thin  for \he\ as well as \h.
As we will show below, this assumption holds only if the power law
distribution of the \h\ forest column densities extends to
$ N << 10^{12}$~\cl.

Previous work either claim that it cannot be determined whether the
observed
\he\ opacity is due to the known forest lines (Jakobsen 1996),
or it can entirely be attributed to them (Songaila et al. 1995).
In this paper we present improved calculations of the line opacity
produced
by the \he\ \lya\ forest.
To overcome the limitations of previous work, we use a Monte-Carlo
technique
to simulate the absorption spectrum of \he\ forest. While the forest-line
interpretation may not describe the physical nature of the IGM, we use it
to
compare with the published results.
Under this scenario, we explore how variations in the minimum column
density,
in the \hep\ to \hi\ population ratio ($\eta$) and in the ratio of helium
to
hydrogen velocity dispersion ($\xi$) affect the
contribution of the known forest components to the total \he\ opacity.
Our calculations yield \he\ opacities smaller
than previously estimated, and we conclude that
forest lines with $N > 2 \times 10^{12}$~\cl\ {\em cannot} account for the
observed \he\ opacity with the parameters commonly assumed.

\section{OPACITY CALCULATIONS}

To evaluate the opacity in single lines we use the standard
curve-of-growth
method with a Voigt profile (see Press \& Rybicki 1993).
Although deviations from Voigt profiles are expected in recent models due
to the spatial extension of the IGM fluctuation, the effect
is not significant for most lines. A typical physical size of such a
structure is about $0.14 h^{-1}$ Mpc (Bi \& Davidsen 1997), equivalent to
14 \kms\ in the velocity space. This is smaller than the
characteristic velocity of $\sim 30$ \kms\ (Press \& Rybicki 1993). The
contributions from density wings at $\sim 100$~\kms\ may become noticeable
only for lines with $N > 10^{14}$~\cl.
Line blending due to thermal and/or
turbulence velocity affect all models, and therefore Voigt profiles are
a fair match at the first order.

The most important factor governing the \he\ opacity is $\eta$.
For very tenuous gas ($N(He II) < < 10^{14}$ \cl), the \he\
absorption decrement is proportional to the optical opacity, but it is not
the case for most
forest lines, due to saturation effects.
For \lya\ lines of equivalent width $W > 0.3$~\AA,
the \he\ to \h\ opacity ratio
for a single feature
changes very slowly with $\eta$ as both \he\ and \h\ lines are
saturated and their equivalent widths depend only weakly on the column
density.

The \h\ and \he\ curves are similar, as \hep\ ions are hydrogenic.
The Doppler parameter for helium depends on $\xi$, which lies between 0.5
and
1.0, depending on the nature of the line broadening.
Observations by Cowie et al. (1995) suggest $\xi \sim 0.8$.
We will use this value in most of our calculations as the results can be
scaled to suit other $\xi$ values.

Assuming $\xi = 0.8$, we show in Fig.~1 the \he\ to \h\ decrement ratio
of a single absorption feature as a function of column density
for three values of the Doppler parameter $b_H$:
20, 30, and 50 \kms.
For a typical value of $b_H = 30$~\kms, this ratio
is near unity for $N \approx 10^{14-15}$~\cl. At higher densities, it
increases as the damping wings contribute appreciably to the \he\
\lya\ absorption. At lower densities, the ratio becomes high as the
\h\ absorption lies on the linear part of the curve and decreases.
{}From Fig.~1 we can immediately infer that the opacity of the \he\ forest
will depend strongly on the minimum column density chosen for the \lya\
forest.

To calculate the wavelength-averaged opacity of the \he\ \lya\ forest, we
start with the standard empirical description of the \h\ \lya\ forest
population. The number of \lya\ forest absorption features, $n$, in quasar
spectra may be described by
\begin{equation}
{{\partial^2n \over \partial z \partial N} = A (1 + z)^\gamma N^{-\beta}}
\end{equation}
(\markcite{Peterson 1978; Tytler 1987; Press \& Rybicki 1993}).
We adopt $\beta = 1.5$ and $\gamma = 2.4$ from \me\ \& Ostriker (1990).
The normalization factor $A$ affects the forest opacity appreciably, and
it
can vary depending on the particular line of sight.
We choose $A$ to be consistent with the number of lines and the measured
\h\ forest opacity for the particular quasar under study.
The Doppler parameter $b$ is assumed to follow a
$\Gamma$-function distribution:
\begin{equation}
p(b)\ db \propto b^{(b_0/b^*) -1} exp(-b/b^*) \ db,
\end{equation}
where $b_0 = 38$~\kms\ and $b^* = 14$~\kms\ (Press \& Rybicki 1993).

Our Monte-Carlo method randomly generates centroid wavelengths,
column densities and Doppler parameters for each line in the \h\ forest
according to Eq.~3 (for a given minimum column density)
and the Doppler-parameter distribution of Eq.~4.
The Voigt profiles of the selected lines are then calculated in the quasar
rest frame with a spacing of 0.01~\AA\ ($\sim 3$ \kms).
The product of these profiles gives the total transmission of the \h\
forest.
The effect of line blending is naturally taken into consideration in this
treatment.
Given the transmission of the \h\ forest we calculate the decrement
$D_{\HI}$
and express the wavelength-averaged forest optical depth as
$\tau_{\HI} = - \ln (1 - D_{\HI})$.
Individual realizations of the forest opacity in this method show
fluctuations of $\sim 15\%$ due to the random line blending.
For each set of parameters we run 10 simulations and average the results
to obtain the final opacity values.

The corresponding \he\ forest is constructed by changing the column
density of each simulated line from $N$ to $\eta N$, and the Doppler
parameter from $b$ to $\xi b$.
The average decrement for \he, $D_{\HeII}$, is calculated,
and the wavelength-averaged optical depth is
given again by $\tau_{\HeII} = - \ln (1 - D_{\HeII})$.

To illustrate our results, we present calculations for a quasar redshift
$\rm z_{em} = 2.74$ for comparison to the HUT observations of HS1700+64
(Davidsen \etal\ 1996).
We normalize the column-density distribution using $A = 2.0 \times 10^7$
to match the line number density measured for this quasar by
Rodr\'{\i}guez-Pascual \etal\ (1995) and Zheng et al. (1997).
We use the 1050--1170~\AA\ wavelength range (in the rest frame)
to calculate the \h\ opacity
and rest-frame wavelengths 262-292~\AA\ for the \he\ opacity.
These wavelength ranges, free of confusion from the
proximity effect and Ly$\beta$ lines, has often been used to calculate the
average absorption decrement $D_A$ (Jenkins \& Ostriker 1991; Press,
Rybicki,
\& Schneider 1993).
Fig.~2 displays the average \he\ forest optical depth as a function of
the minimum column density for two values of redshift.
The effect of changing $\eta$ is shown by
several curves in the upper panels of the figure.
The lower panels show the sensitivity of the results to the assumed value
of $\xi$ with $\eta$ held fixed at 100.

The behavior of the \he\ forest opacity shown in Fig. 2 can be
qualitatively understood in terms of the \he\ to \h\ decrement ratios
illustrated in Fig. 1.
The total \h\ opacity increases as the minimum column density is lowered,
but it levels out at $N \approx 10^{13}$~\cl.
This is because at low column densities \h\ absorption is
proportional to the column density and becomes insignificant
as lines move off the saturated portion of the curve of growth.
The \he\ opacity, on the other hand, continues to increase significantly
as the minimum column density is lowered toward $N \approx 10^{11}$~\cl.
This is because \he\ absorption is still in the flat part
of the curve of growth. The opacity ratio $\rho$ therefore becomes higher
when more weak forest lines are included.
Note that strong lines ($ N > 2 \times 10^{13}$~\cl) produce nearly the
same
amount of absorption for \he\ and \h, but they contribute only $\sim 30\%$
of the total \he\ opacity for $\eta =100$.
Weak lines ($2 \times 10^{12}~\cl < N < 2 \times 10^{13}$~\cl)
contribute $\sim 40\%$ of the total \he\ opacity, compared with only
$\sim 5\%$ for \h.

We expect the forest opacity to increase with redshift in proportion to
the
number of \h\ forest clouds within the wavelength interval of interest.
Since $\Delta N \propto (1 + z)^\gamma \Delta z$, and
$\Delta z = (1 + z) \Delta \lambda_o$,
where $\Delta \lambda_o$ is the rest frame wavelength interval,
the \he\ forest opacity should have a redshift-dependence of
$(1+z)^{1+\gamma}$. To verify this we have also made calculations
applicable
to the quasar Q0302-003.
To match the wavelength range used by Hogan \etal\ (1997) to measure the
\he\ opacity in their GHRS spectrum, we use the rest-frame wavelength
ranges of
1157-1195~\AA\ and 289-299 \AA\ for the \h\ and \he\ forest, respectively.
We normalize the column density distribution to give a total \h\ forest
opacity of 0.21 (see below). These results are shown in
Fig. 2$b$.

Given opacities for both the \he\ and \h\ \lya\ forest, we can calculate
the ratio of their average opacities, $\rho$, as a function of the minimum
column density. This ratio is shown in Fig.~3 for $z_{em} = 2.74$.
(A comparison to $\rho$ calculated for $z_{em} = 3.3$ indicates that the
behavior of $\rho$ is virtually independent of redshift over this range.)
Qualitatively this is to be expected, following the derivation of $\rho$
in the optically thin limit in Eq.~3.
However, this relation assumes that the decrement produced by an
absorption
line is proportional to its central optical depth.
This is only correct for optically thin lines
(i.e., a centroid optical depth not significantly greater than unity) and
hence may not be applied to strong forest lines, which account for
$\sim 60\%$ of the \h\ optical opacity.
At column densities $N \approx 10^{14}$~\cl, for example, the value
of $\rho$ is nearly unity even for $\eta > > 1$.
As shown in Fig.~3, only at column densities $N < < 10^{12}$~\cl\ does
$\rho$ approach the value predicted by Eq.~2
(assuming a reasonably high value of $\eta > 50$).
Fig.~3 also shows that applying Eq.~2 at
$N_{min} \approx 2 \times 10^{12}$~\cl\
leads to a considerable overestimation of the \he\ forest opacity.
\newpage
\section{COMPARISON TO PREVIOUS WORK}

For comparison to previous work on the quasar Q0302-003, we refer to
the \he\ opacities shown in Fig. 2$b$.
For parameters typical of those used by Jakobsen \etal\ (1994),
Madau \& Meiksin (1994) and Giroux et al. (1995), i.e. $\eta = 100$,
$b_H = 30$~\kms, $\xi = 0.5$, and $N > 10^{12}$~\cl,
we obtain an optical depth for the \he\ forest $\tau_{\HeII} \approx
0.53$.
Their calculated opacities are, approximately, 0.9, 1.3 and 1.1,
respectively.
These values are larger than ours by $\sim 70 - 150 \%$.
A difference at a level of $\sim 15\%$ is due to line blending.
Weak lines, when blended with strong ones, do not decrease the
transmission
appreciably in already blank regions of the spectrum.
We estimate that, at $\rm z \approx 3$, $\sim 15\%$ of the spectral
region is virtually blank due to the high opacity of strong lines,
implying
that at least $\sim 15\%$ of the \he\ opacity from weak lines would not
be realized.
The rest of the discrepancy may be attributed to the normalization chosen
for
the distribution of \h\ \lya\ forest lines (Eq.~3). A high normalization
leads
to a high \h\ opacity as well as a high \he\ opacity.
The average \h\ optical depth in the spectrum of quasar Q0302-003 is
particularly low, with $\tau_{\HI} \approx 0.21$ between 5000 and
5200~\AA\
as estimated from Fig.~1 of Songaila \etal\ (1995). This is mainly due to
a void in this quasar spectrum (Dobrzycki \& Bechtold 1991; Nath \& Sethi
1996).
For comparison, the \h\ forest opacities for the line distributions used
by
Jakobsen et al. (1994), Madau \& Meiksin (1994) and Giroux et al. (1995)
are
0.35, 0.55 and 0.44, respectively.

An even larger discrepancy exists between our result and that of Songaila
\etal\ (1995) who simulated a \he\ spectrum based on their Keck spectrum
of the quasar Q0302-003.
Their simulated \he\ spectrum based on the assumed values of $\eta = 80$
and $\xi = 1.0$ produces a very high value of $\tau_{\HeII} \approx 1.6$,
corresponding to $\rho \approx 5$.
As we showed in \S 2, however, saturation in the \he\ \lya\ lines prevents
$\rho$ from reaching values this high unless the minimum column density
is significantly lower than $ 10^{12}$~\cl.  Our simulations using the
same
parameters yield $\tau_{\HeII} \approx 0.9$, i.e. $\rho \approx 3$.

The large difference between our result and that of Songaila \etal\
(1995),
on the order of $\Delta\tau_{\HeII} \approx 0.7$, probably arises from
several factors:
(1) noise can contribute significantly to the
 \he\ opacity in their simulated spectrum.
 $\eta = 80$ represents an optical depth ratio of 20. At a noise level of
 2\% (S/N = 50 per resolution element), an amplification by 20 may produce
 an erroneous optical depth of $\sim 0.4$.
(2) Even at an infinite S/N level, the \he\ spectrum simulated by
 Songaila \etal\ (1995) contains not only the contribution from lines with
 $ N > 2 \times 10^{12}$~\cl, but also those below this threshold that
 have not been identified.
 Some very broad but shallow lines have been seen in the \lya\ forest
(Kirkman
 \& Tytler 1995), and they may be from clouds with $N < 2 \times
 10^{12}$~\cl.
(3) Metal lines are treated as additional forest lines.
 Observations have shown that metal lines often account for more than
5\%
 of the total absorption lines in the \lya\ forest (Carswell \etal\ 1991;
 Rauch \etal\ 1993).
 In some quasars such as HS1700+64 (Rodr\'{\i}guez-Pascual \etal\ 1995)
 the density of metal lines in the \lya\ forest region is about 20\% of
the
 true \lya\ forest lines. When ``amplifying'' the whole \h\ forest
spectrum
 these metal lines may add a significant amount of erroneous opacity to
the
 simulated \he\ spectrum.

To avoid these problems we constructed a \he\ spectrum
using the fitted \lya\ forest lines in Q0302-003 tabulated by
Hu et al. (1995) instead of the real spectrum.
In the optical range of 4400--5000~\AA, Hu et al. (1995) identified 258
forest lines with $N \geq 2 \times 10^{12}$~\cl.
Using their $b$ and $N$ values we construct an
\h\ spectrum and another for \he, assuming $\eta = 80$ and $\xi = 1.0$.
Because the reconstructed spectrum contains neither noise, nor weak
features
of $N < 10^{12}$~\cl, nor metal lines,
it is different from the real spectrum in some aspects.
For example, $\sim 60\%$ of the data points in our reconstructed \h\
spectrum
have optical depths larger than 0.05, in comparison with 75\% in the real
spectrum (Songaila \etal\ 1995).
{}From our simulated spectrum we calculate the average \he\
optical depth in the $D_A$ region to be $\sim 0.91$.
This value is significantly smaller than that derived by Songaila
\etal\ from the {\em same} spectrum, but in good agreement with the
predictions of our Monte-Carlo technique. Our simulated spectrum is
plotted
in Fig.~4 and can be compared with Fig.~1 of Songaila \etal\ (1995).

\section{DISCUSSION}

Until recently, most studies of the \h\ \lya\ forest were limited to
features
with column densities exceeding $\sim 10^{13}$~\cl.
Fig. 2$a$ shows that, for this limiting column, the \he\ forest can
contribute
no more than 50\% of the total observed opacity even for $\xi = 1.0$ and
$\eta$ as high as 1000.
Observations of \h\ \lya\ features with an order of magnitude lower column
density in other quasars with the Keck 10-m telescope
(Songaila \etal\ 1995; Hu et al. 1995) suggest that the power-law
distribution
of the \h\ \lya\ forest may extend to column densities as low as
$ 2 \times 10^{12}$~\cl.
Even at this limiting column, however, matching the \he\ opacity observed
with HUT requires $\eta > 1000$.
The quasar Q0302-003 has been observed with Keck to a limiting column
density of $ 2 \times 10^{12}$~\cl.
Fig. 2$b$ shows that at this limiting column density $\eta >> 1000$ is
also
required to match the observed \he\ opacity $\tau_{He II} \approx 2.0$
(Hogan et al. 1997).

Is it possible that the metagalactic background radiation
is so soft that $\eta > 1000$?  We think it is unlikely.
Theoretical calculations of the UV-background radiation produced by the
observed quasar population, taking into account filtering
by the \lya\ forest, predict that $\eta$ should lie in the range 30 to 100
(\me\ \& Ostriker 1990; Madau \& Meiksin 1994; Haardt \& Madau 1996).
A mean quasar spectrum (Zheng et al. 1997$a$) suggest a continuum shape of
$f_\nu \approx \nu^{-1.8}$ between 1 and 4 Rydberg, and, if the ionizing
field
is dominated by quasar radiation, this would lead to $\eta \simeq 100$
(Madau, private communication).
Background radiation dominated at low energies by a large contribution
from
young galaxies could produce $\eta > 1000$ (\me\ \& Ostriker 1990).
Several observational details argue against $\eta$ being this high,
however.
First, very soft background radiation would
produce a significant proximity profile (Zheng \& Davidsen 1995), which
is not observed.
Second, if the observed \he\ opacity is derived entirely from forest lines
with $N > 2 \times 10^{12}$~\cl,
there should be spectral windows in which the \he\ opacity is much lower.
In these windows (``voids") there are no or very few forest lines detected
in the Keck spectra of Q0302-003. The recent observations with improved
resolution (Hogan et al. 1997) show a substantial \he\ opacity
even in the spectral voids.
The lack of emission peaks corresponding to these voids in the wavelengths
shortward of the redshifted \he\ \lya\ emission suggests that the
contribution
to the total opacity from observed forest lines is limited.
Third, the inferred shape of the background ionizing spectrum based on
observations of metal lines in \lya\ clouds suggests that $\eta < 155$
(Songaila et al. 1995).

If the background radiation is hard enough to ensure $\eta \sim 100$,
then the additional opacity required to explain the observations of
HS1700+64 and Q0302-003 must arise from  a more tenuous part of the
IGM. If the minimum column density can be extended to arbitrarily low
values so that the formula of \me\ (1994) applies, then the HUT
observations
of $\tau_{\HeII} = 1.00 \pm 0.07$ in HS1700+64 and the \h\ opacity of 0.22
(Davidsen \etal\ 1996) imply $\rho = 4.5 \pm 0.5$.
For $\xi = 0.8$, the \he\ to \h\ population ratio must be
$\eta = 100_{-20}^{+25}$. Similarly, for the quasar Q0302-003, we have
$\tau_{\HeII} \approx 2.0$ and $\tau_{\HI} = 0.21$ imply $\rho \gtsim 10$
and
$\eta > 180$. Hogan et al. (1997) find that the \he\ optical depth
in the spectral voids of Q0302-003 is $\sim 1.3$.
In \S~3 we estimate  the \he\ opacity produced by the
forest lines (Hu et al. 1995) $\tau_{He II}\sim 0.9$. The sum of these two
terms (2.2) is  very close to the measured total \he\ opacity (2.0),
suggesting that our estimate of the forest opacity is quite accurate.

Croft et al. (1997) use gravitation-instability calculations to simulate
the \he\ absorption observed by HUT. Their results confirm that much of
the opacity arises in underdense regions in which the \h\ opacity is
very low. It is true that the extension of Eq. 3 to column densities far
below $10^{12}$ \cl\ will indeed make forest-line model indistinguishable
from the gas-dynamics model.
However, the Keck spectrum of HS1700+64 (Zheng \etal\ 1997$b$) suggests
that,
at $\rm z \sim 2.4$, the number of lines with $N < 10^{13}$~\cl\ is
considerably smaller than the extrapolation of a power law predicts.
This trend is qualitatively consistent with the cold-dark-matter models
(\me\ et al. 1996; Bi \& Davidsen 1997) that the distribution of forest
lines
flattens at lower column densities.

\section{SUMMARY}

Previous work overestimated the contribution from known forest lines.
Assuming a column-density distribution for the \h\ forest with a
power-law index $\beta = 1.5$, a
\hep\ to \hi\ population ratio $\eta = 100$ and velocity dispersion ratio
$\xi = 0.8$, we calculate a \he\ opacity produced by the observed forest
lines that is $\sim 3 \times$ the \h\ opacity.
This can account for about half of the observed \he\ opacity
along the sight line to the quasars HS1700+64. For Q0302-003, the lower
limit
to the \he\ opacity set by the observation is at least $8 \times$ the \h\
opacity, making it more difficult to explain in terms of forest lines.
Our results show that about 50\% of the \he\ opacity at $\rm z \sim 3$
is produced by the most tenuous part of the IGM with
column density $ N < 10^{12}$~\cl.

The detection of \he\ absorption concludes a long pursuit and opens
a new window to the early universe. Because $\tau(\HeII)/ \tau(\HI) =
\eta/4$
and $\eta >>1$,
\he\ absorption spectra are sensitive to the underdense region where the
early
seeds of density fluctuations develop. With future far-UV spectroscopy
at higher resolution and sensitivity, we will be able to resolve
individual
structures along the full paths to the early universe, measure the
ionization
state, the density, and distribution of baryonic matter in intergalactic
space.

\acknowledgments

This work has been supported by NASA contract NAS-5-27000 to the Johns
Hopkins University and grant AR-05821.01-94A from the
Space Telescope Science Institute, which is operated by the Association of
Universities for Research in Astronomy, Inc., under NASA contract
NAS5-26555.

\clearpage

\begin{figure}
\caption{Ratio of \he\ and \h\ \lya\ decrement vs. column density of
neutral hydrogen for a single line. 
$\xi = b_{He}/b_H = 0.8$ is assumed.
Dotted curve: $b_H = 20$~\kms, solid line: $b_H = 30$~\kms, and
dashed curve: $b_H = 50$~\kms.}
\end{figure}

\begin{figure}
\caption{Wavelength-averaged \he\ optical depth produced by the \h\ forest
lines with column density between the minimum value and $10^{17}$~\cl.
The upper panels hold $\xi$ fixed at 0.8 while varying $\eta$, and the
lower panels hold $\eta$ fixed at 100 while varying $\xi$ as shown.
The values at the current observation limit, $\log N = 12.3$
are marked, representing the maximum amounts of opacity the observed
forest lines can account for. For Figure~2$a$,
the quasar redshift is assumed to be 2.74, and the opacity is calculated
for
the $D_A$ region, i.e. between 262 and 292~\AA\ in the quasar rest frame.
The observed \he\ opacity, $\tau_{\HI} = 1.00 \pm 0.07$, in the HUT
observation of HS1700+64 is marked by the shaded area. For Figure~$2b$,
the quasar redshift is assumed to be 3.3, and the opacity is calculated
between 292 and 303.3~\AA\ in the quasar rest frame, assuming an average
\h\
opacity $\tau_{\HI} = 0.21$, as estimated from the work of Songaila \etal\
(1995). The shaded region is set by the recent results $\tau_{\HeII} =
2.0^{+1.0}_{-0.5}$ (Hogan et al. 1997).
}
\end{figure}

\begin{figure}
\caption{Ratio of $\rho$ He~{\sc ii} forest opacity to H~{\sc i} forest
opacity
vs. minimum column density.
A redshift $z_{em} = 2.74$ is assumed, but as
discussed in the text, $\rho$ is virtually independent of redshift.
The upper panel holds $\xi$ fixed at 0.8 while varying $\eta$, and the
lower panel holds $\eta$ fixed at 100 while varying $\xi$ as shown.
The values at the current observation limit, $\log N = 12.3$
are marked.
The dashed lines represent the expected values derived using
Eq.~2 with the same parameters used to generate the curves.
}
\end{figure}

\begin{figure}
\caption{Simulated spectrum of quasar Q0302-003. Upper panel: hydrogen
forest absorption spectrum, constructed from the fitted lines with
column density
$N > 10^{12}\ \rm cm^{-2}$~ in the Keck spectrum (Hu et al. 1995). Lower
panel:
He~{\sc ii} forest spectrum, constructed by multiplying the optical depth
at each pixel by 20 and dividing the wavelengths by 4.
The pixel size is 0.04~\AA\
for the H~{\sc i} spectrum and 0.01~\AA\ for He~{\sc ii}.
The He~{\sc ii} spectrum is binned to 2.5~\AA\ for display.
It shows an average decrement of 0.60, representing an optical depth of
0.91,
which is significantly smaller than that derived by Songaila \etal\
(1995).
Note that the wavelength range used to calculate the $\tau_{\HeII} \approx
2.0$ corresponds to 4960--5120~\AA, where the \h\ opacity is even lower.
}
\end{figure}

\clearpage
\setcounter{figure}{0}
\begin{figure}
\plotone{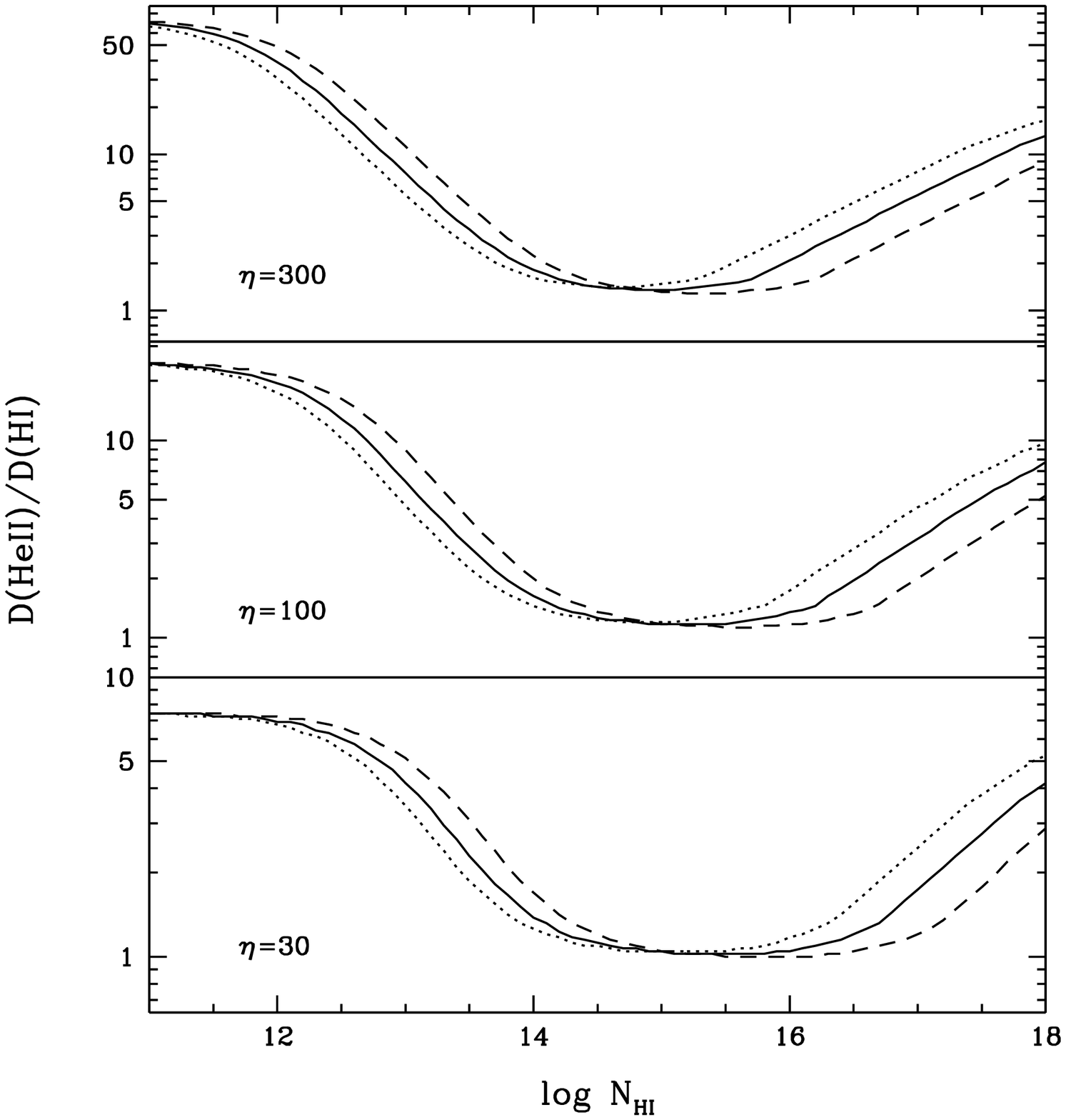}
\caption{~~}
\end{figure}

\begin{figure}
\plotone{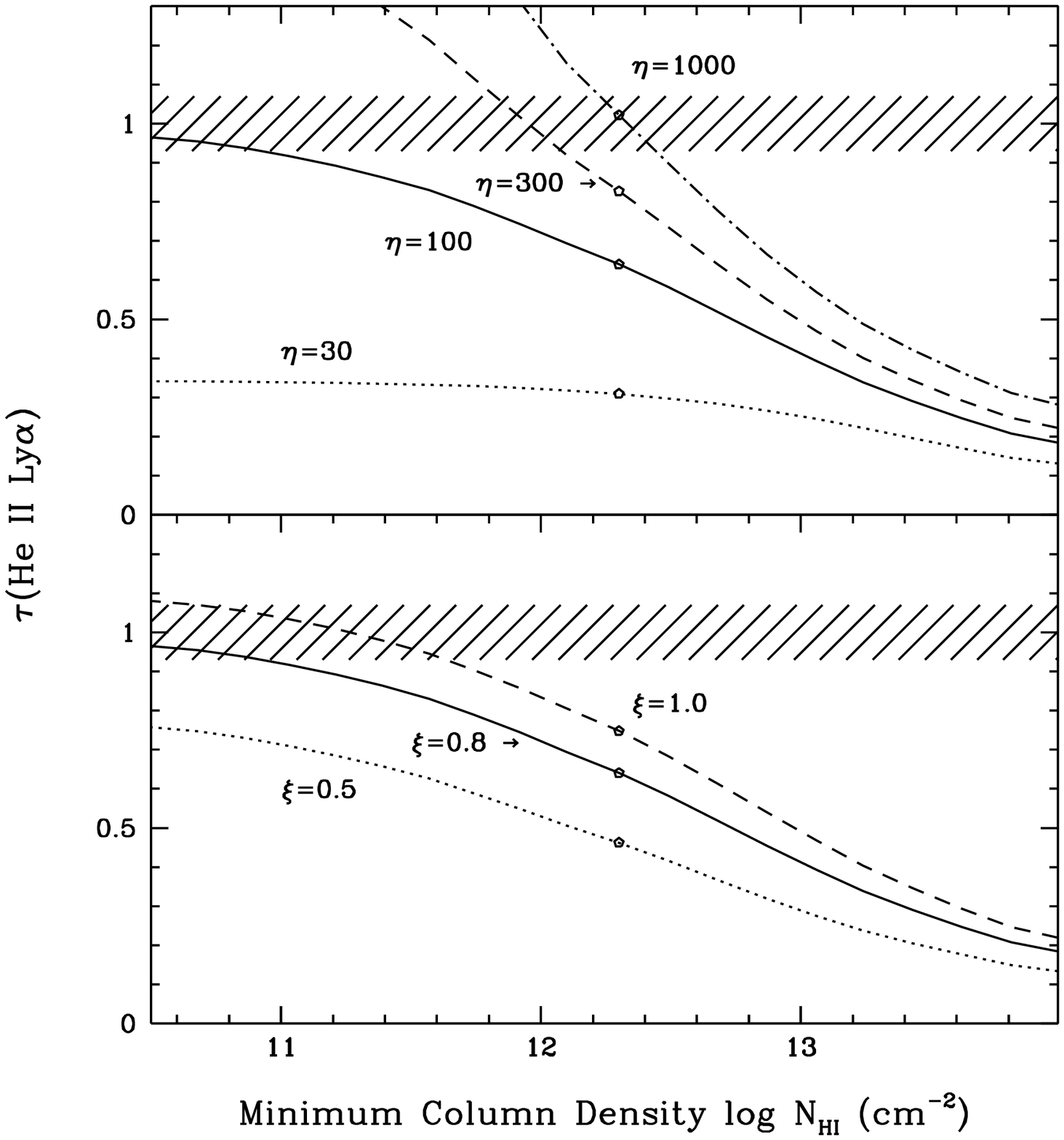}
\caption{~~}
\end{figure}

\addtocounter{figure}{-1}
\begin{figure}
\plotone{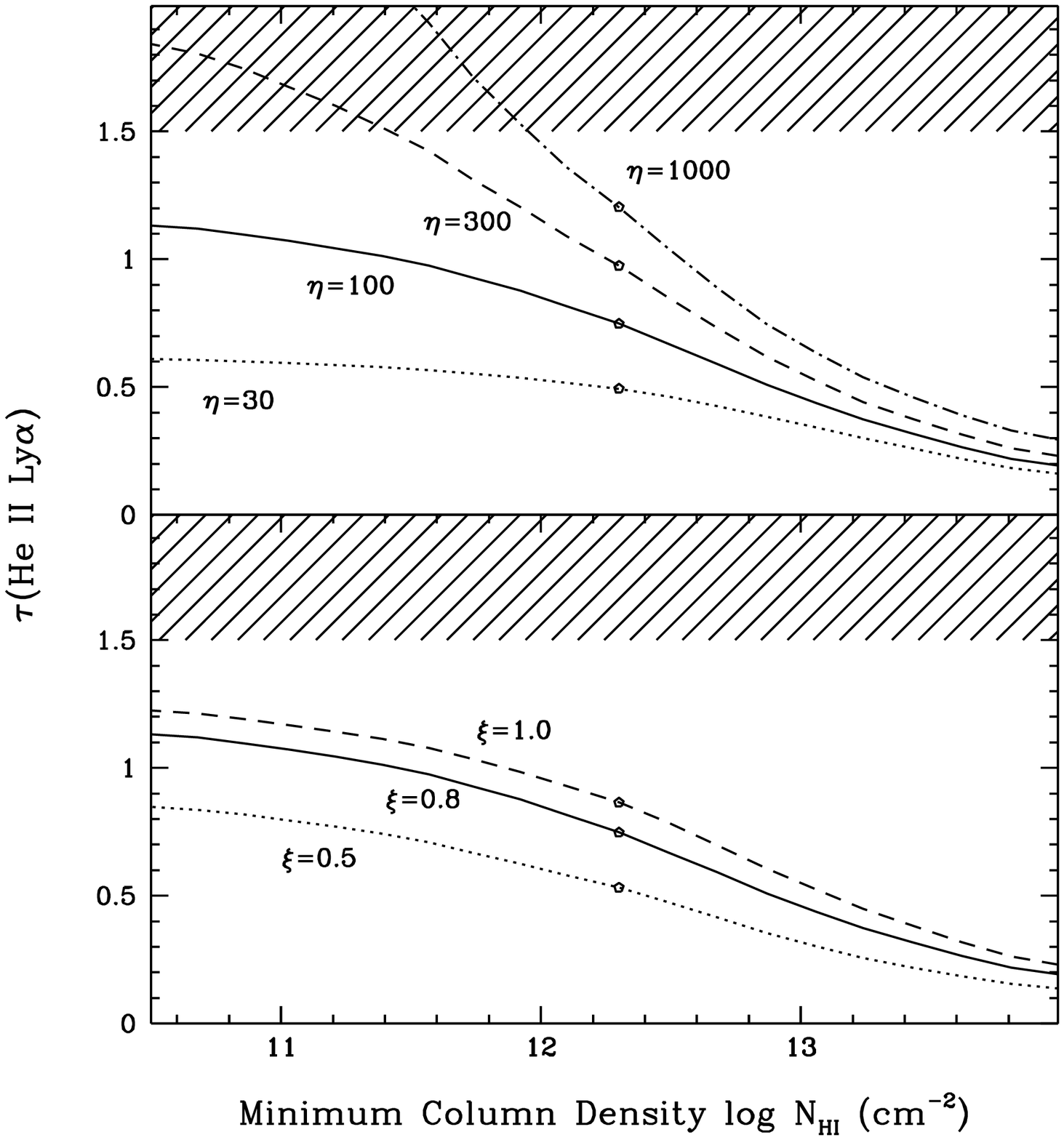}
\caption{continued}
\end{figure}

\begin{figure}
\plotone{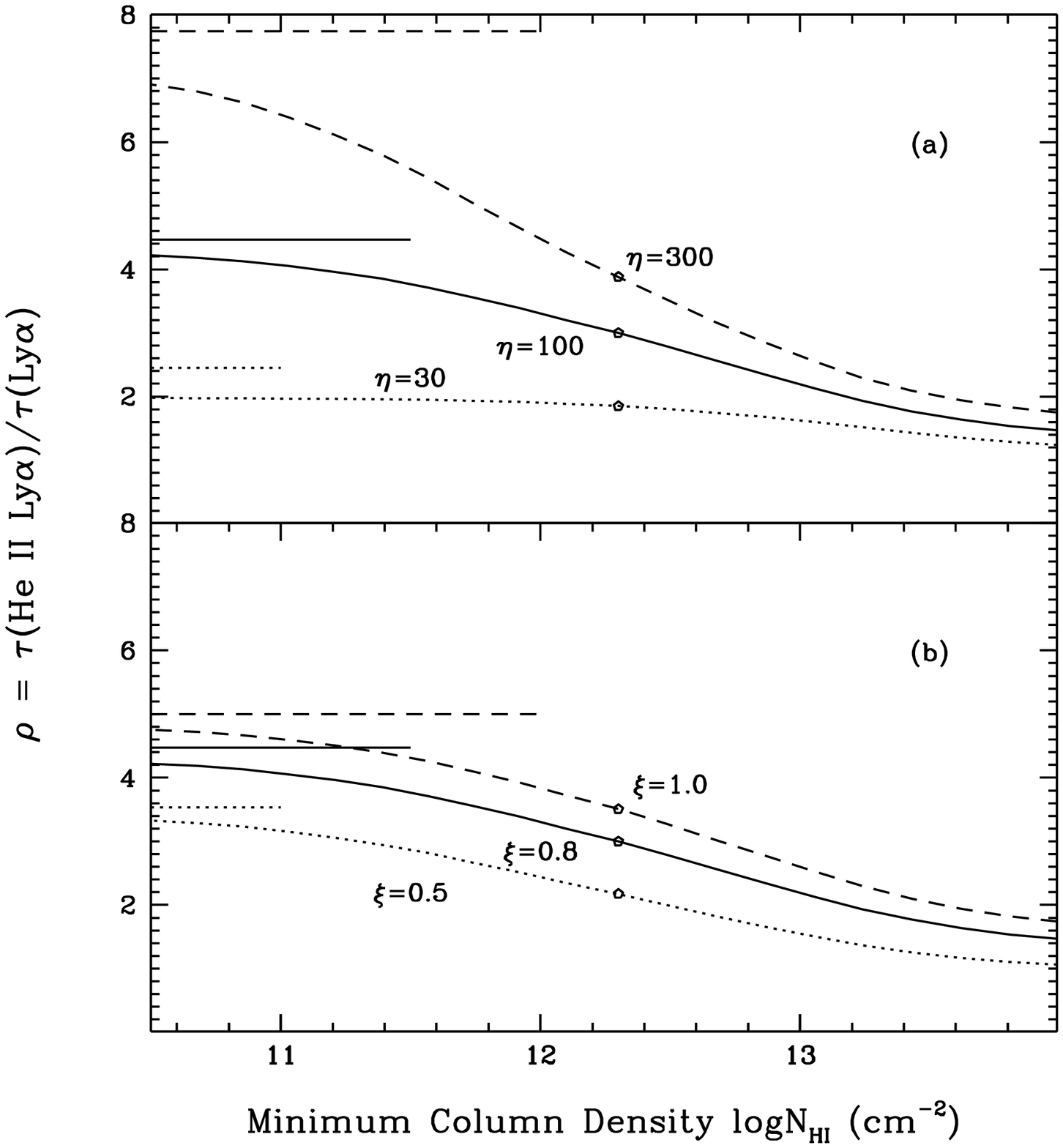}
\caption{~~}
\end{figure}

\begin{figure}
\plotone{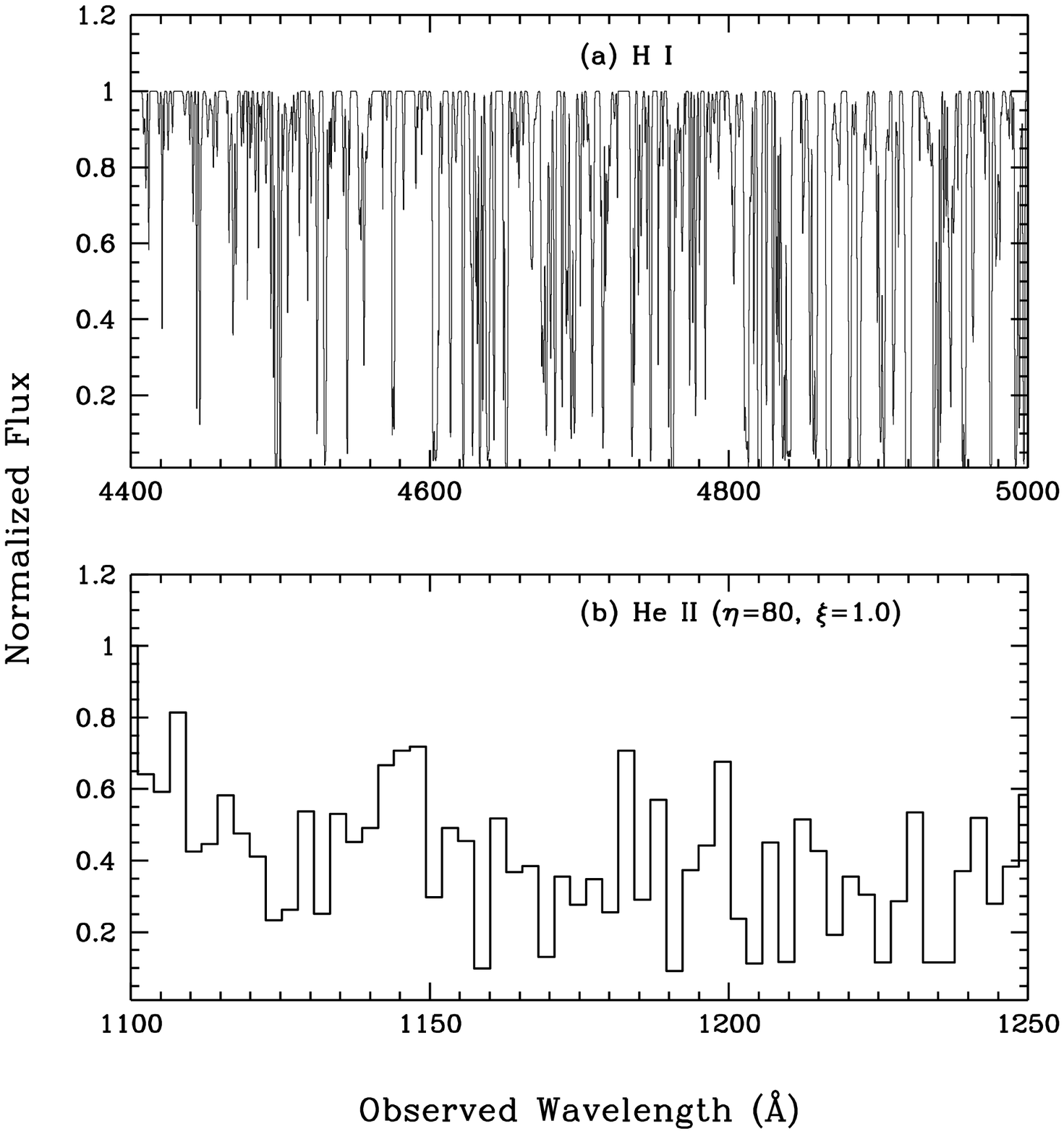}
\caption{~~}
\end{figure}

\end{document}